\documentclass{nature}
\usepackage{graphicx}
\usepackage{ccaption}
\usepackage{caption}
\usepackage{placeins} 
\usepackage{float}

\usepackage[hyphens]{url}
\usepackage[hidelinks]{hyperref} 
\hypersetup{breaklinks=true}
\makeatletter 
\let\saved@includegraphics\includegraphics
\AtBeginDocument{\let\includegraphics\saved@includegraphics}
\renewenvironment*{figure}{\@float{figure}}{\end@float}
\makeatother  

\usepackage{lineno} 

\title{Drag Reduction of a Circular Cylinder Through the Use of an Architectured Lattice Material}

\author{M. Pelacci$^{1}$, A. G. Robins$^{2}$ \& S. Szyniszewski$^{3}$}
\begin{document}

\maketitle

\begin{affiliations}
	\item University of Surrey, Dept. of Civil and Environmental Engineering
	\item University of Surrey, Dept. of Mechanical Engineering, EnFLo Lab
	\item University of Durham, Dept. of Engineering
\end{affiliations}

\begin{abstract}
\section*{Abstract}
Materials with periodic architectures exhibit many beneficial characteristics such as high specific stiffness thanks to the material placement along the stress paths and the nano-scale strength amplification achieved through the use of hierarchical architectures. Recently, the porosity of architectured materials was leveraged to increase the efficiency of compact heat exchangers, and their internal aerodynamics was studied. However, their performance on external aerodynamics applications is generally assumed to be detrimental. Here, we demonstrate that exposing 3D lattice material to the external flow reduced the drag of a circular cylinder when placed at carefully selected angular locations. We tested two configurations with the lattice material installed at the windward and leeward regions. On the one hand, the windward configuration showed a strong Re dependency, with a drag reduction of up to 45\% at Re=11E4. On the other hand, the lattice material in the leeward region reduced the drag by 25\% with weak Re dependency. Alterations of the lattice material topology had a noticeable effect on the drag reduction in both cases. Adding aerodynamic features to the already proven beneficial structural properties of 3D lattice materials might aid in the development of low-powered automotive, naval, and aerospace vehicles. 
\end{abstract}

\section*{Introduction}
Innovative manufacturing techniques such as 3D printing and 3D weaving facilitate the development of lattice materials with unprecedented control over their architectures down to the nano-scale\cite{valdevit_protocols_2011,liu_dynamic_2014,pham_damage-tolerant_2019,queheillalt_cellular_2005,wadley_multifunctional_2006,zheng_multiscale_2016}. These multiscale, hierarchical structures exhibit unprecedented specific stiffness because they leverage the material properties at the nano-scale\cite{schaedler_ultralight_2011}. In addition to the structural benefits, recent studies have also taken advantage of their multiphysical properties such as damping\cite{ryan_damping_2015,salari-sharif_damping_2018} and thermal conductivity\cite{maloney_multifunctional_2012,zhao_permeability_2014}. Human-made lattice materials have made possible compact heat exchangers exploiting the combination of internal flow, increased surface area, and thermal conductivity enabled through the use of metallic base material\cite{zhao_experimental_2016,zhao_combining_2017,siemens,welch_micro-lattice_2020}. 

Observation of natural systems indicates that porous media can also modify the aerodynamics of external flows. Tree canopies can be viewed as permeable structures, which affect environmental flow and mixing, with associated effects on temperature distribution, mixing of water vapour, and CO2 absorption\cite{belcher_wind_2012}. At a smaller scale, dandelion seeds have a porous structure, which enhances their ability to disperse. Wings have a hierarchical porous architecture comprising feathers that are made up of tiny barbules interlinked by thin membranes. These ease the wing actuation by behaving as unidirectional valves that block the air flow when the wing moves downwards, and open when the wing moves upwards. Noticeably, the barbules spacing is independent of the wings’ size, and  remains practically constant\cite{sullivan_scaling_2019} at 8-16 $\mu m$ over a wide range of  bird sizes, ranging from the Anna's hummingbird, 6 g, to the Andean condor, 10 kg. Similarly, an owl’s wings have spatially distributed feathers that reduce the aerodynamic noise generated during flight. Man-made structures, such as stochastic metallic foams, have also recently been employed for aerodynamic noise suppression\cite{bae_effect_2011}. In addition to noise generation, aerodynamic drag is another fundamental issue in engineering because it affects material usage in infrastructure components, ranging from light-posts to high-rise buildings, and the energy consumption of bluff bodies such as cars and trucks, hence affecting climate change.

While passive devices, such as riblets\cite{kim_drag_2016,bechert_experiments_1997}, have been developed to reduce friction drag on streamlined bodies such as aerofoils, alternative solutions are employed for bluff shapes, where the drag is pressure dominated and is highly dependent on the location of the flow separation from the surface. For example, dimples\cite{bearman_control_1993,bearman_golf_1976,wong_means_1979} and grooves\cite{lim_flow_2002,kimura_fluid_1991} over spherical or cylindrical surfaces are established solutions for delaying flow separation, thus reducing drag. Local and distributed surface protrusions have also been investigated, in the form of tripping devices\cite{guven_surface-roughness_1980,pearcy_flow_1982, ekmekci_control_2011} or random surface roughness (e.g. using sandpaper)\cite{guven_surface-roughness_1980, achenbach_influence_1971, zdravkovich_effect_2003}. Similar mechanisms can be found in nature, for instance, in the form of spanwise grooves on Cacti plants\cite{talley_flow_2002} or distributed roughness such as the plumage of bird wings. Recently, some initial attempts have been made to reduce drag via the application of porous coatings. Possibility for the drag reduction through the use of porous materials was first suggested computationally\cite{bruneau_numerical_2008, bruneau_passive_2004, bruneau_control_2006, bruneau_coupling_2010,bruneau_active_2012,bruneau_effect_2014}. However, wrapping an isotropic porous material around the whole circumference of a circular cylinder did not delay flow separation, but increased the overall drag\cite{naito_numerical_2012,showkat_ali_bluff_2016}. On the contrary, the application of porous inserts at the rear side of a circular cylinder reduced the drag\cite{klausmann_drag_2017} by 13\%, while simultaneously reducing the total weight of the structure.

Here, we investigate the effect of three-dimensional (3D) architectured porous material on the external flow around a circular cylinder. We 3D-printed and tested a total of four material topologies (Fig. \ref{Fig:ExperimentalSetup}a-b), starting with the architecture inspired by previous optimisation studies of structural\cite{zhang_fabrication_2015} and thermal\cite{zhao_experimental_2016} properties, from now on termed the Default topology. We placed the material over the Leeward region, extended to the area where the flow separates (Fig. \ref{Fig:ExperimentalSetup}c) or over the Windward region, where the positive pressure load is experienced (Fig. \ref{Fig:ExperimentalSetup}d).

We measured the total drag applied over the span of a smooth cylinder, and the drag force of our configurations employing 3D architectured porous coatings (Fig. \ref{Fig:ExperimentalSetup}e, f). The static pressure was also acquired at the mid-span (Fig. \ref{Fig:ExperimentalSetup}g) to gather information about the effect of the porous substrate on the mean flow. Further, static pressure measurements were acquired at both the solid, inner surface and at the external surface of the porous coating in order to gain further understanding of the flow field within the porous substrate. Based on our insights from the first phase of tests, we designed and manufactured additional topologies to improve the overall fluidic performance, and to assess the sensitivity of the flow to changes in the 3D porous material architecture.

\begin{figure}
	\centering
	\includegraphics[width=0.99\textwidth]{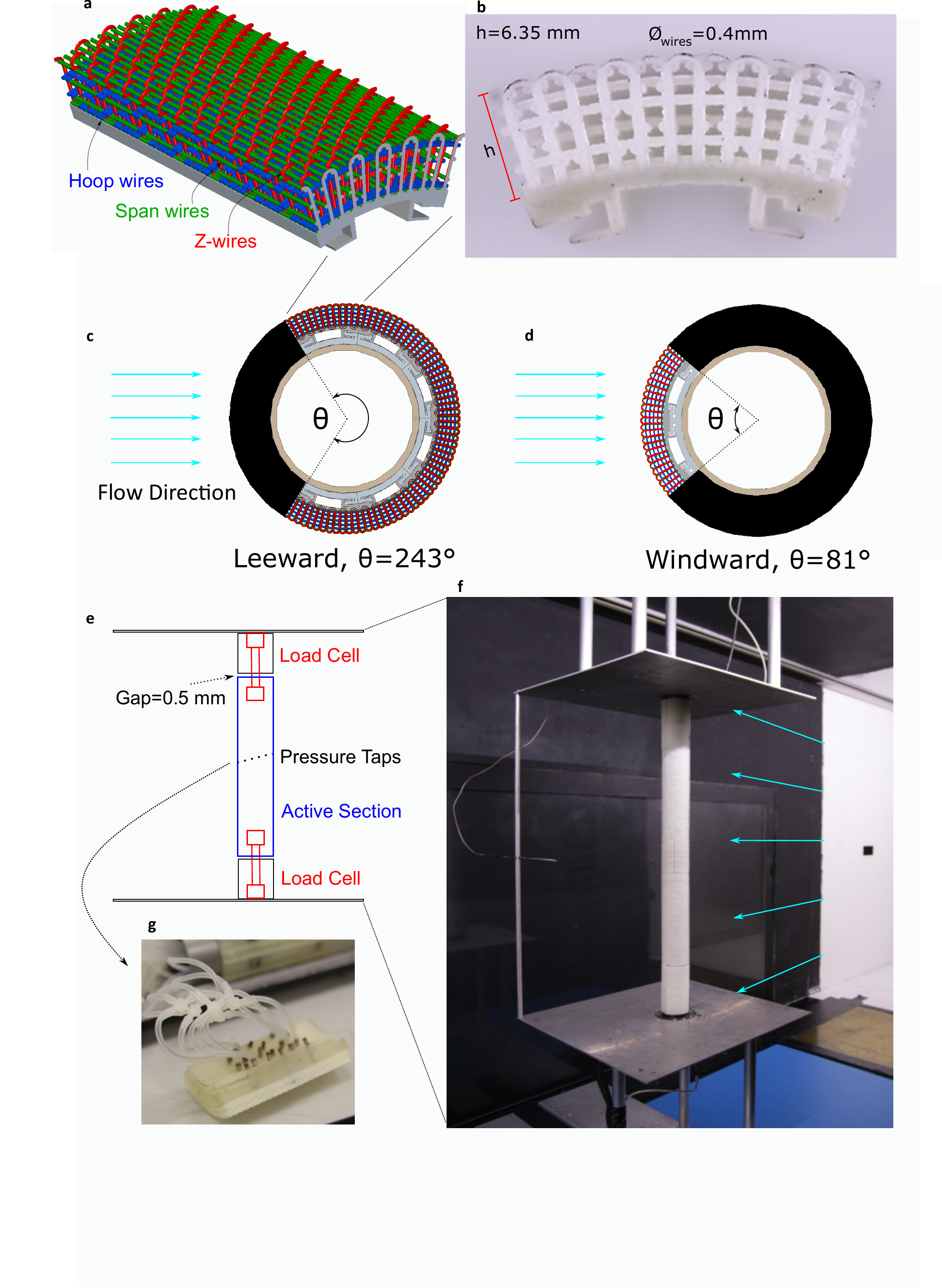}
	\caption{\textbf{\rule[-2pt]{1.5pt}{13.5pt} \hspace{1pt} The wind-tunnel study of the architectured materials for drag reduction of circular cylinders.}}
\end{figure}
\begin{figure}[h]
	\centering
	\contcaption{\textbf{\rule[-2pt]{1.5pt}{13.5pt}} \hspace{1pt}\textbf{a,} the architectured lattice consists of hoop wires (blue), Z-wires (red) and spanwise wires (green). \textbf{b,} the 3D printed module used for the aerodynamic tests. The wire diameter is 0.4 mm and the module thickness 6.35 mm. \textbf{c-d,} the lattice sheath extended over 243 degrees for the leeward and 81 degrees for the windward cases. The configurations were assembled from 20 to 60 3D-printed modules to cover the required locations. \textbf{e,} experimental model split into an active section mounted to the load cells, which is encompassed by the shorter cylindrical segments. The active section of 65\% of the cylinder's total span guarantees that the drag measurement is not affected by the boundary layer developed over the end-plates. \textbf{f,} the static pressure readings were taken at mid-span through the use of T-junctions to average the pressure readings inside the porous medium. \textbf{g,} the assembled experimental model of the leeward configuration during wind-tunnel testing.}
	\label{Fig:ExperimentalSetup}
\end{figure}

\FloatBarrier

\section*{Architectured porous inserts over the leeward region}
The coating at the Leeward region reduced the drag coefficient, $C_D$, by 15\% compared to the smooth case (Fig. \ref{Fig:LeewardPlacement}a) and by 35\% compared to the configuration with an empty void instead of the porous inserts, termed the Slotted case. Analysis of the pressure coefficient, $C_p$ profiles at the lowest and the highest Reynolds numbers tested (Fig. 2b-c) showed an enhanced recovery of the base pressure coefficient, $C_{pb}$, for the porous case compared with the smooth configuration. This was further enhanced when the porous case was compared against the Slotted case (Fig. \ref{Fig:LeewardPlacement}b-c), whose geometry favoured flow separation precisely after the step edge, flattening the $C_p$ profile. This fact indicates that recovery is facilitated by the downstream shift of the flow detachment point. To maintain an attached flow over the step, the coating needs to exert passive suction on the fluid, drawing it towards the inner surface of the cylinder. In other words, a radial velocity component in equation (\ref{Eq:NS}) produces a local curvature of the mean streamline towards the wall. This radial component generates a centrifugal pressure on the underlying fluid, which balances the measured pressure gradient (Fig. \ref{Fig:LeewardPlacement}d) with the convective and the Reynolds stress terms in the Navier-Stokes equations for momentum transport. At the location where the internal and the external $C_p$ profiles intersect, the mean streamline straightens, and flow separation commences. The occurrence of separation deflects the streamline outwards from the coating, producing a radial pressure gradient of the opposite sign. The disappearance of the radial pressure gradient is the signature of the flow separation from the surface. The inferred qualitative flow behaviour is illustrated in (Fig. \ref{Fig:LeewardPlacement}e). For further analysis of the Navier-Stokes equations, we refer to the Methods section (Fig.\ref{Fig:NavierStokes}).
\begin{equation} \label{eq:NS_momentum_radial}
	u_r\frac{\partial u_r}{\partial r} + \frac{u_\theta}{r}\frac{\partial u_r}{\partial \theta} -\frac{u^{2}_\theta}{r} = -\frac{1}{\rho}\frac{\partial p}{\partial r} - \left(\frac{1}{r}\frac{\partial \tilde{u}_\theta \tilde{u}_r}{\partial \theta}+\frac{1}{r}\frac{\partial (r\tilde{u}^{2}_r)}{\partial r}-\frac{\tilde{u}_\theta^{2}}{r}\right)+\nu\left(\nabla^{2}u\right)+F_r
	\label{Eq:NS}
\end{equation} 
\begin{figure}
\centering
	\includegraphics[width=0.8\textwidth]{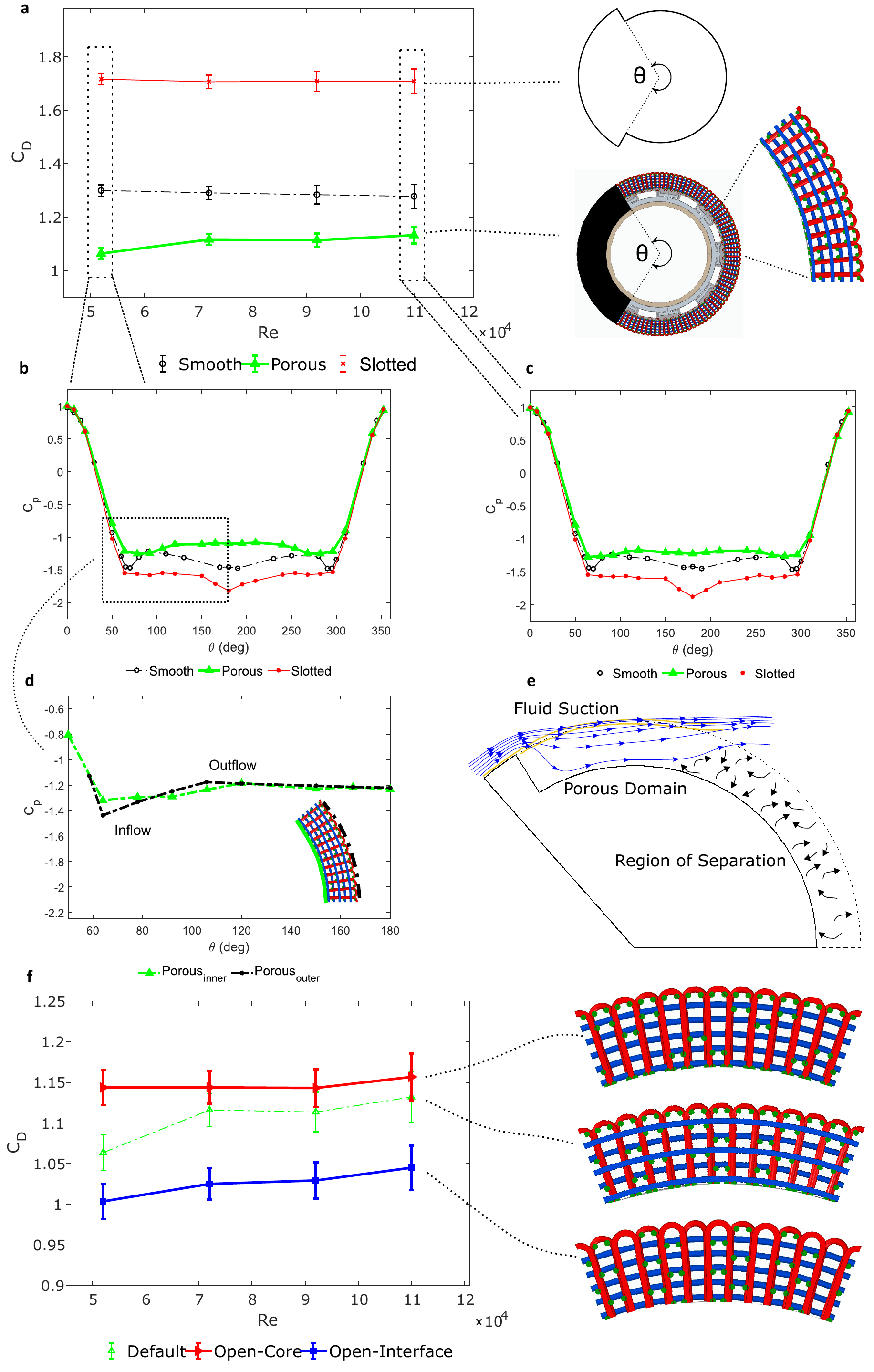}
	\caption{\textbf{\rule[-2pt]{1.5pt}{13.5pt} \hspace{1pt}The effect on the drag coefficient of the architectured material at the leeward location.}}
\end{figure}
\begin{figure}[h]
	\contcaption{\textbf{\rule[-2pt]{1.5pt}{13.5pt}} \hspace{1pt}\textbf{a,} the drag coefficient reduced by 15\% and exhibits a weak dependence on Reynolds number (Re), with the higher drag reduction at the lower Re. Conversely, a slotted configuration, without a porous layer, showed 30\% drag increase compared to the smooth cylinder.  \textbf{b-c} the enhanced base pressure recovery reduced the drag, indicating attached flow over the portion of the porous material. Conversely, the absence of pressure recovery for the slotted configuration implies flow separation at the steps. \textbf{d,} static pressure measurements at the internal and external porous surfaces were consistent with inflow and outflow regions along the porous medium. \textbf{e,}  measurements indicate passive suction from the architectured porous medium, which averts the flow separation at the step. \textbf{f,} modified lattice topology at the interface increases transport between the external flow and the intra-lattice, secondary flow, which reduces the drag by an additional 6\%. The results confirm that the external flow is significantly sensitive to the internal topology of the material.}
	\label{Fig:LeewardPlacement}
\end{figure}
We inferred that passive suction was a distinctive feature of placing the coating at the Leeward region, regardless of its topology and was influenced by both the architectured material cover angle, 243$^{\circ}$ in our test, and the material topology itself. Since our primary goal was to investigate the effect of the material architecture on the external flow, we kept the coating coverage region constant and only manipulated the material architecture. We considered the fluid-porous interface as the most relevant region for cross-flow momentum transport, similarly to forest canopies. Firstly, we modified the initial topology by increasing the porosity of the external layer, resulting in the Open-Interface topology, in order to improve the supply of energetic fluid across the fluid-porous interface. The Open-Interface design reduced $C_D$ by 24\% compared with the 15\% achieved by the Default architecture, maintaining a weak dependency on Re, \textbf{Fig. \ref{Fig:LeewardPlacement}f}. Secondly, we also investigated to what extent the performance was dependent on the internal flow. We tested another material topology with a reduced number of wires in the core, termed as the Open-Core configuration. It produced the lowest drag decrease of the three architectures tested, with the average $C_D$ reduction of 12\% compared with the 15\% of the Default case. Nevertheless, the significant differences between the results confirmed the influence of the layout of the internal wires on the aerodynamic performance \textbf{(Fig. \ref{Fig:LeewardPlacement}f)}. While at the lowest range the Default topology produced a $C_D$  4.5\% lower than that of the Open-Core, the effect of the core architecture diminished, and discrepancies fell within the experimental uncertainties at the higher range of Re. This observation indicates that 3D lattice architecture needs to be optimized not only in the context of the overall body geometry, but also the working flow regime, namely the Re number range.

\section*{Architectured porous inserts at the Windward region}
The Windward configuration reveals a different working mechanism from the previously discussed Leeward configuration. $C_D$ decreased sharply with Re, a reduction ranging from 12\% at the lower Re, to 40\% at the highest compared with the smooth cylinder case \textbf{(Fig. 3a)}. We also compared the performance of 3D architectured lattice material inserts against the smooth cylinder equipped with trip-wires of d/D=0.9\%, placed at the same angular locations as the porous coating boundaries. Trip-wires produced a similar trend of drag reduction as did the architectured porous insert in the windward position, but with slightly better efficiency at higher Re. The $C_p$ profiles (Fig. 3b-c) confirmed the trend of $C_D$, displaying a recovery of the base pressure coefficient, coupled with a downstream shift of the separation point as Re increase. As the promotion of turbulence within the boundary layer is a well-established feature of plain cylinders with tripping devices, the similarity of the profiles indicates a turbulent boundary layer over the porous configuration as well. While trip-wires lead to ‘kinks’ in the $C_p$ profiles \textbf{(Fig. 3b-c)} for the generation of local separation bubbles, these are not present when the porous coating is in place, suggesting an alternative mechanism of turbulence promotion within the boundary layer.
\newline
The architectured porous substrate modifies the $C_p$ distribution over the region $0^{\circ} < \theta < 40.5^{\circ}$ compared with the smooth cylinder. Remarkably, the pressure coefficient is lower than unity at the stagnation point, and generally lower than that of the smooth case for $0^{\circ} < \theta <30^{\circ}$ (Fig. 3d). The material location favours fluid penetration into the porous medium, with associated dissipative losses due to friction, resulting in a drop in $C_p$ and a radial pressure gradient directed outwards through the porous substrate. This indicates that the inlet region extends between $0^{\circ} < \theta < 30^{\circ}$, the pressure gradient switches direction. Then, at $\theta=30^{\circ}$, the external flow penetration ceases (Fig. 3e), and the flow continues by ‘skimming’ the top region of the porous substrate. Considering continuity, fluid cannot accumulate inside the porous region and must be expelled from the coating, with the outlet region likely distributed over $30^{\circ} < \theta < 40.5^{\circ}$, where secondary flow mixes with the external, primary flow, in fashion similar to that seen in distributed, passive jets (Fig. 3e). These perturb the primary flow, eventually, promoting turbulence. Based on these insights, we designed a third topology, with internal 'channels' following the expected streamlines of the secondary, intra-lattice flow to maximize the flow rate of the distributed passive-jets, which is called the Passive-Jets case. We removed the interface wires over the inferred inlet at the stagnation region with the rationale of facilitating the penetration of the incoming fluid. The internal architecture was redesigned to minimize the internal flow losses in the azimuthal direction, and the outlet region was designed towards the end of the porous coating, by promoting outflow in accordance with the previous analysis.
\begin{figure}
	\centering
	\includegraphics[width=0.8\textwidth]{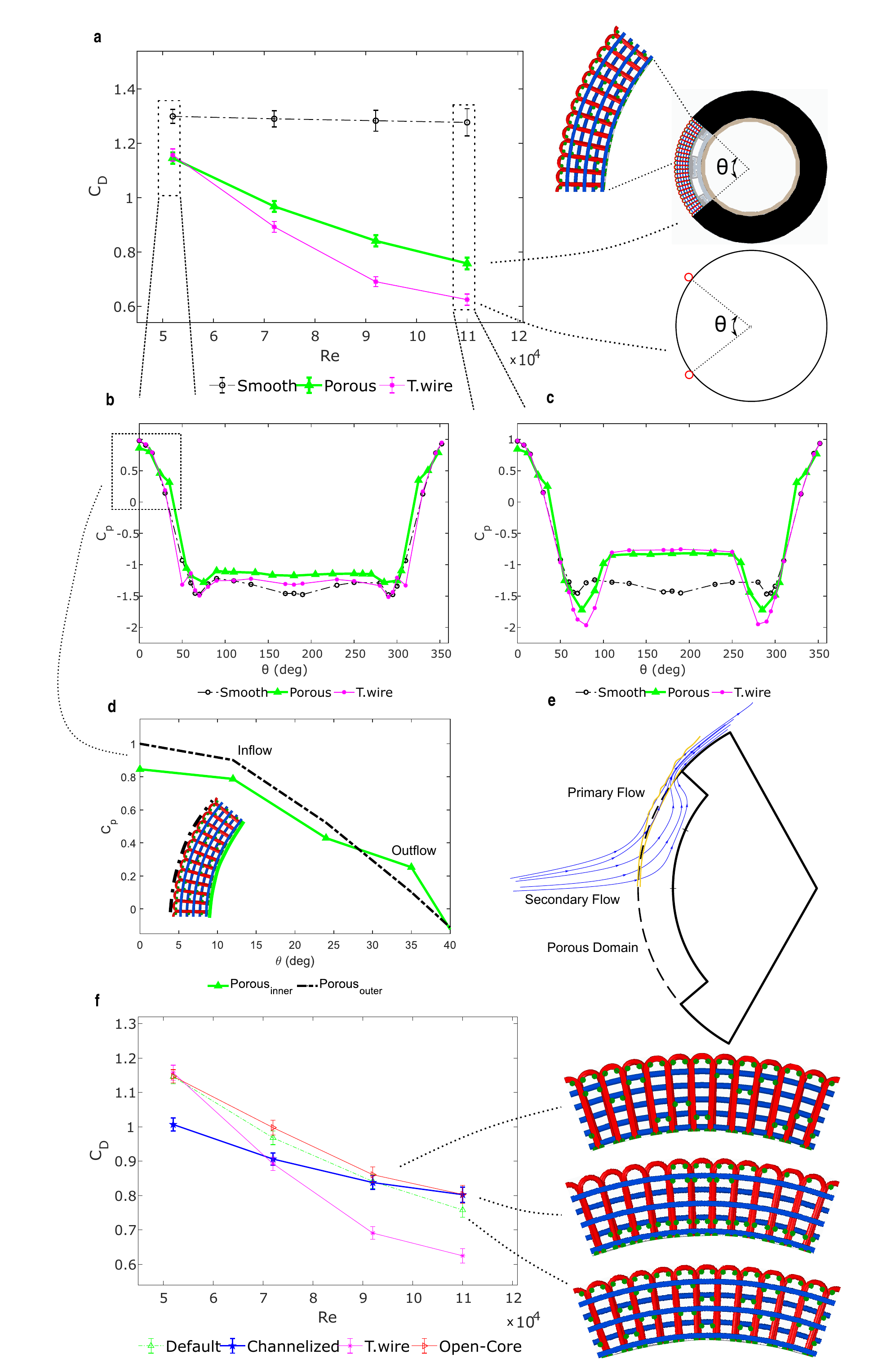}
	\caption{\textbf{\rule[-2pt]{1pt}{13.5pt} \hspace{1pt} The effect on the drag coefficient of the architectured material at the windward location.}}
	\label{Fig:WindWardPlacement}
\end{figure}
\begin{figure}[h]
	\contcaption{\textbf{\rule[-2pt]{1.5pt}{13.5pt}} \hspace{1pt}\textbf{a},  the drag coefficient exhibits a strong dependence on the Re, with the lowest drag at high Re. We compared the performance with trip-wires, which are conventionally used to trigger turbulence in the boundary layer. \textbf{b-c,} pressure measurements reveal the increase of the base pressure, consistent with turbulent transition in the boundary layer, which shifts the flow separation point downstream, thus reducing the drag. The absence of kinks at the $\theta = 50^{\circ}$ locations in the configuration with the lattice material points to a different mechanism for turbulent transition in comparison with the trip wire. \textbf{d,} static pressure measurements at the internal and external porous surfaces are consistent with inflow and outflow regions along the porous medium. \textbf{e,} measurements suggest the development of a secondary flow inside the medium, which interacts with the external flow near the outlet from the secondary flow at ca. the $\theta = 35^{\circ}$ region. \textbf{f,} modified lattice topology favoured the development of secondary flow along the envisaged streamlines. The architecture outperformed the trip-wire at the lowest Re, and maintained the drag reduction similar to the initial topology at the highest Re.}
\end{figure}

Noticeably, the new design outperformed the trip-wire by 13\% at the lowest Re range, and exhibited lower drag than the Default topology up to Re = 90,000. Despite the greater $C_{pb}$ recovery of the Passive-Jets \textbf{(Fig. \ref{Fig:PressureProfileChannelized})}, the pressure profile did not exhibit any kinks over the whole Re number range, in contrast with the smooth cylinder equipped with tripping devices, indirectly confirming the working hypothesis. Despite the promising results at lower Re, the mechanism requires further fine-tuning to avoid the deterioration in performance at the upper range of Re tested. If the intra-lattice flow rate is not sized in accordance with the external, primary flow rate, the passive-jets might thicken the boundary layer. Consequently, the flow might separate early, leading to a reduced recovery of $C_{pb}$, and thus a smaller drag reduction. Active intra-lattice permeability control, demonstrated in active filters\cite{krautz_structural_2015}, could enable a wider operational range in the future.

\FloatBarrier
\section*{Discussion}
Certainly, the impact of the flow angle of attack on the drag reduction and on $C_D(Re)$ is of primary relevance in these experiments. The Windward configuration outperforms the Leeward one in terms of the drag reduction magnitude by 33\% at the upper end of the Re range, with the same lattice topology in both cases \textbf{(figs. 2a, 3a)}. However, the Leeward configuration displays an almost constant $C_D(Re)$ trend against Re, and ensures that the Windward configuration exhibits the best performance at the lower range of Re tested. The results of the Channelized topology \textbf{(fig. 3f)} for the Windward configuration and those of the Open-Interface topology \textbf{(fig. 2f)} for the Leeward configurations further confirm the above trends, and the relevance of the flow angle of attack on the $C_D$ reduction and on the $C_D(Re)$ trend. The modification of the angle of attack even biases the effect of the topology modification as per the comparison of the Open-Core topology results for the two configurations \textbf{(figs. 2f, 3f)}. Remarkably, the Windward configuration shows minimal differences in $C_D$ between the Open-Core, and the Default topologies, while the Leeward cases reveal differences between 2\% and 7\% depending on Re. Indeed, the topology must be tuned according to angular location around the cylinder to ensure an optimal drag reduction under well defined flow conditions. Besides, the analysis proves the important impact of the inner material topology, and the resulting inner flow, on the aerodynamic performance. However, the results of the Channelized and the Open-Interface hint that the impact on performance is weakened as we move further into the porous coating, thus leaving the outer topology as predominant. Further studies focused on the flow within the porous medium are needed to understand, and ultimately control the link between the material angular location, the topology and performance.
\newline
As for effectiveness, 3D architectured porous materials provide a greater reduction over a broader Re range than random surface roughness such as sandpaper, while outperforming selected tripping devices at $Re<65000$ (Fig. 4a). Although a dimpled surface leads to a greater drag reduction than the application of porous inserts, 3D architectured porous substrates exhibited immense flexibility in their performance through modification of their topologies. Splitter plates lead to an equivalent drag reduction at the cost of severe encumbrance, which would be impractical and could pose safety risks if applied to road vehicles. Similarly, cylinders with slots that allow flow penetration and subsequent near-wake bleed or ventilation produce better performance at the cost of structural integrity (Fig. 4b). Nonetheless, further understanding of the aerodynamics of porous media could further improve the flexibility of performance, both in terms of Re effects and the magnitude of CD reductions. These, coupled with the multi-domain applications of 3D architectured porous materials, may leverage their use in the development of sustainable technologies.

The decrease of carbon footprint is a world-wide concern, and heavy-weight vehicles, such as those employed for essential goods transportation, produce a substantial fraction of global carbon emissions that is hard to mitigate\cite{noauthor_regulation_nodate}. Although the geometry of cargo trucks have been modified over the years in favour of rounded shapes, trucks are still affected by pressure drag over extended areas at the front, and at the rear part of the trailer, where local separation occurs. Our results clearly state that the presence of 3D architectured porous materials either at the Windward or the Leeward locations improves performance over rounded, smooth shapes by delaying flow separation. In particular, the Windward configuration could be advantageous around the front area of the cab, for instance at the vertical front pillars, which can be a source of drag when side-wind occurs\cite{minelli_numerical_2016}, and promotes flow separation, which cannot be prevented either by rounded geometries or by passive surface modifications at smaller scales. Similarly, the Leeward configuration could serve as an example to promote the application of porous stripes at the rear edges of the trailer and the cab, as an alternative to flapping appendices, boat tails or splitter plates, all of which involve major geometry modifications. Possibly, porous stripes could redirect the mean flow towards the base region as an additional form of passive jets increasing the base pressure, thus reducing total drag.

Importantly, transport vehicles operate in environments with highly variable, turbulent flow conditions. They experience complex flows resulting from the interaction with other vehicles, and with cross-wind, which modify the nominal flow angle of attack (Fig. 4c-d), and create sudden wind gusts. Thus, the vehicle components such as the front pillars can experience windward (Fig. 4c) or leeward flow (Fig. 4d), depending on the incoming flow conditions. Whereas flow facing exposure would call for dimples, the leeward location would benefit from a splitter plate or slots enabling partial base bleeding. These are radically different modifications. On the contrary, our 3D lattice material system reduced the drag at both the leeward (Fig. 2) and the windward (Fig. 3) regions, even though via distinct mechanisms. Whereas surface roughness or splitter plates work well for special cases, our 3D lattice architected material offers a promise of general solutions under variable and multi-directional flow conditions. The locations calling for the 3D porous components coincide with load carrying frame components, which offers an associated opportunity for structural weight reduction. Our 3D lattice architected porous material could be 3D printed on metallic surfaces directly to merge fluidic and mechanical features. Also, 3D porous materials are beneficial for shock and impact dissipation\cite{smith_steel_2012, szyniszewski_mechanical_2014, tavares_mechanical_2020} during accidents (see Supplementary Video 1).

\begin{figure}
	\centering
	\includegraphics[width=1.01\textwidth]{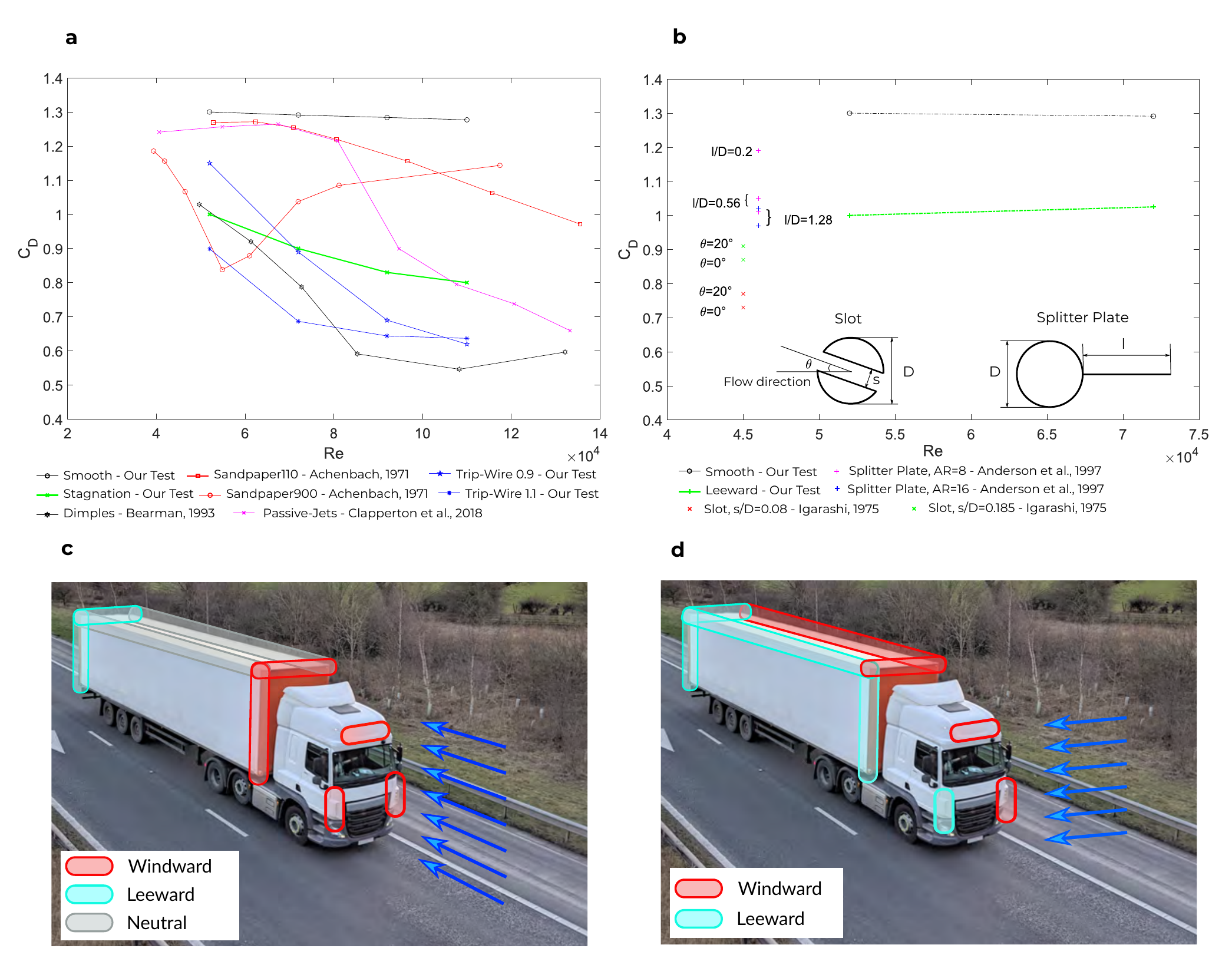}[t]
	\caption{\textbf{\rule[-2pt]{1pt}{13.5pt} \hspace{1pt} Comparison of our work with prior approaches to drag reduction. We also show a potential benefit of 3DW versatility to reduce aerodynamic drag of trucks under variable wind and turbulence conditions.} \textbf{a}, comparison of the Windward configuration against stochastic roughness, trip wires and dimples. \textbf{b,} comparison of the Leeward configuration and other approaches to wake-interference such as splitter plates and slots. \textbf{c, d,} reduce energy consumption of heavy-goods vehicles could be achieved by placing architectured porous coating at their windward and leeward regions; the default material configuration introduced in this study reduced the drag in both windward and leeward actions. }
	\label{Fig:Benchmarking}
\end{figure}

\FloatBarrier


The architectured porous inserts could be tuned to work on a range of large road vehicles and further optimized depending on the mean flow angle of attack or multi-objective scenario-based approaches to obtain the best performance under high-variability flow conditions. Alternatively, active control of the lattice topologies could allow the intra-lattice topology to be adjusted in response to changes and directional variability of the incoming flow. This could be performed by controlling wire topology via dedicated solutions such as electromagnetic fields (see \textbf{Fig. \ref{Fig:MagneticYarns}}), use of shape memory alloys, via internal fluid pressure if hollow wires were employed, or mechanical loading in the case of bi-stable buckling configurations. 

Reduction of the side force and, perhaps more importantly the rolling moment in crossflow is a challenge that architectured porous structures can address, leading to improved safety and operability. The aim would be to interfere positively with the flow field over the upper part of the trailer side face and roof, in particular to weaken the vortex system that develops over the roof in yawed flow and decrease the overall side force. Only devices capable of acting effectively in flows from any direction can achieve this. Optimisation of architectured porous structures for this purpose is likely to be a complex task, given the range of flow conditions and geometries of interest, but one clearly worth pursuing. The multi-parameter design of these structures offers a broad range of potential applications, such as those discussed above, but simultaneously makes the search for optimum configurations a difficult process.

Active flow control could be implemented via dynamic changes of the intra-lattice topology using morphing architectures and flow sensing solutions. These could be achieved through the use of magnetic stimuli applied to functional yarns with embedded magnetic particles\cite{krautz_structural_2015, krautz_new_2015, zimmermann_mechanics_2014, krautz_hysteretic_2017}, as demonstrated in \textbf{Fig. \ref{Fig:MagneticYarns}} in the Methods section. Alternative approaches to active control are also possible, such as bi-stable buckling lattice architectures, which switch between two configurations under prescribed in-plane loading or lattice geometries controlled by insertion of fluidic actuators as building elements. In addition to controlling the material configuration based on the incoming flow conditions, active control could promote the use of porous media for both drag reduction and drag generation systems. For instance, one may need to minimise the drag during the acceleration, or the cruise velocity phase, and to maximise it during the deceleration phases. The material topologies could also be adjusted for vehicles operating in swarm configurations.

The reduction of aerodynamic drag is by no means the only ground transportation industry's concern. There is intensive ongoing research on heat transfer for cooling applications, especially in the context of electric vehicles where water cooling of the battery is not practical due to the risk of electrical shock to passengers. The advantages of porous substrates over flat surfaces for heat transfer have been demonstrated over recent years\cite{mahjoob_synthesis_2008}. Although the implication of complex geometries still impedes a thorough  understanding of the physics\cite{woudberg_geometric_2012}, the studies highlighted the importance of the internal convection on the heat transfer efficiency of porous media when high Re flows are involved, hence the value of the internal flow for thermal performance. This feature might broaden the applicability of 3D porous material, when applied over thermal sources such as the front region of a truck's cab where the engine is located or over the battery compartment of electric vehicles. 3D lattice structures were tested as compact heat exchangers\cite{zhao_combining_2017} for wall-bounded flows; hence they lend themselves as good candidates for merging the drag reduction and the heat transfer capabilities in the context of external vehicle aerodynamics. Thus, our 3D architectured porous technology could function as both drag reduction and heat exchanger device.

Nevertheless, future work is needed to accomplish the afore-listed goals. Firstly, to emulate the flow around trucks' cabs or trailers, or other automotive applications, 3D lattice substrates needs to be tested at a Re range potentially beyond $10^6$. Secondly, the attainment of the maximum drag reduction requires an understanding of the physics of the transport across the porous surfaces linking the intra-lattice with the vehicle length scales, followed by optimisation studies of the parameters characterising the material topology such as the locations of the wires along the span-wise, the cross-flow and the stream-wise directions, the relative diameters of the lattices and gap sizes, as well as the thickness of the inserts, and, ultimately, their locations over the body. The tuning of a large number of design parameters could enable the achievement of multiple objectives, such as reduced drag, a preferred range of vortex shedding frequencies, and a desired rate of internal heat transfer, while satisfying the weight and the structural requirements. This will inevitably require that an accurate model of the flow fields and any structural or heat transfer interactions is built so that a true optimisation process can be undertaken. At the same time, increased empirical understanding of the issues involved will be an essential ingredient in making progress.

\newpage
\section*{Methods}

Although the analysis cannot be conclusive due to the lack of velocity measurements within the medium, it helps to provide a qualitative picture of the flow pattern. By assuming the zero value at the wall of the velocity and the Reynolds stresses, it is possible to analyse the sign of each term in the equation, with the exception of the Reynolds shear stresses. The first term is responsible for the convection of the fluid within the medium, hence of curving the mean streamlines and generating the centrifugal term, also acting to balance the negative pressure gradient. The second term cannot be fully determined as the azimuthal distribution of the radial velocity is not known a priori. It must start at zero at the step edge and decrease up to an unknown location over the medium, contributing to the negative pressure gradient before increasing towards positive values. However, the convective terms are not the only ones responsible for balancing the negative pressure gradient, as proved by the contribution of the Reynolds stresses. The analysis suggests the depicted streamlines pattern over the first portion of the porous medium, which generate passive suction on the fluid. 

In contrast, over the second portion of the substrate, the internal fluid must be expelled once it separates from the wall, and the associated streamlines must be curved outwards. This gives rise to the centrifugal term, which is claimed to be mainly responsible for balancing the positive pressure gradient. In fact, all the remaining terms are capable of balancing a negative pressure gradient only, except the second convective term and that associated with the Reynolds shear stresses. While the former can play a role over the portion of the coating where the azimuthal distribution of the radial velocity decreases only, the latter cannot be determined, and the reliable measurements or predictions of the internal flow are needed. However, it is not necessary to provide a qualitative description of the impact of the porous substrate on the mean streamline path. It must be noted that the dissipative term does not play a relevant role in balancing the pressure gradient, suggesting that the drag reduction working mechanism is distinct from streamlining.

\begin{figure}[H]
	\centering
	\includegraphics[width=0.80\textwidth]{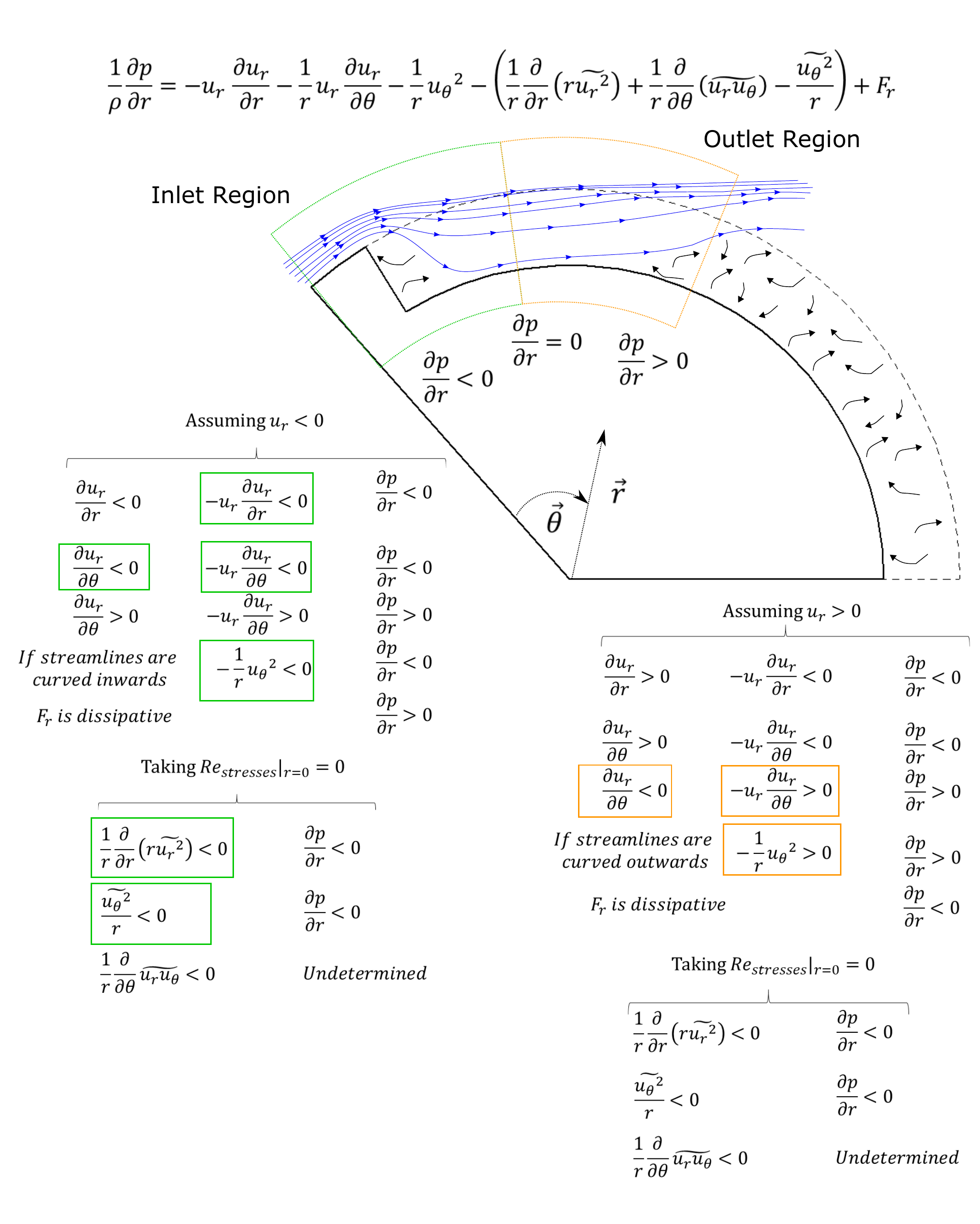}
	\caption{\textbf{\rule[-2pt]{1pt}{13.5pt} \hspace{1pt} The analysis of the inner, and the outer static pressure measurements with the aid of the Navier-Stokes equations.} The figure shows the inlet (green) and the outlet (orange) flow regions of the Leeward configuration. Maintaining an attached flow over the step edge requires the convection of fluid within the substrate, thus $ur<0$. This fact is in agreement with the analysis of the Navier-Stokes equation coupled with the measured pressure gradient in the radial direction. }
	\label{Fig:NavierStokes}
\end{figure}

\begin{figure}[H]
	\centering
	\includegraphics[width=0.85\textwidth]{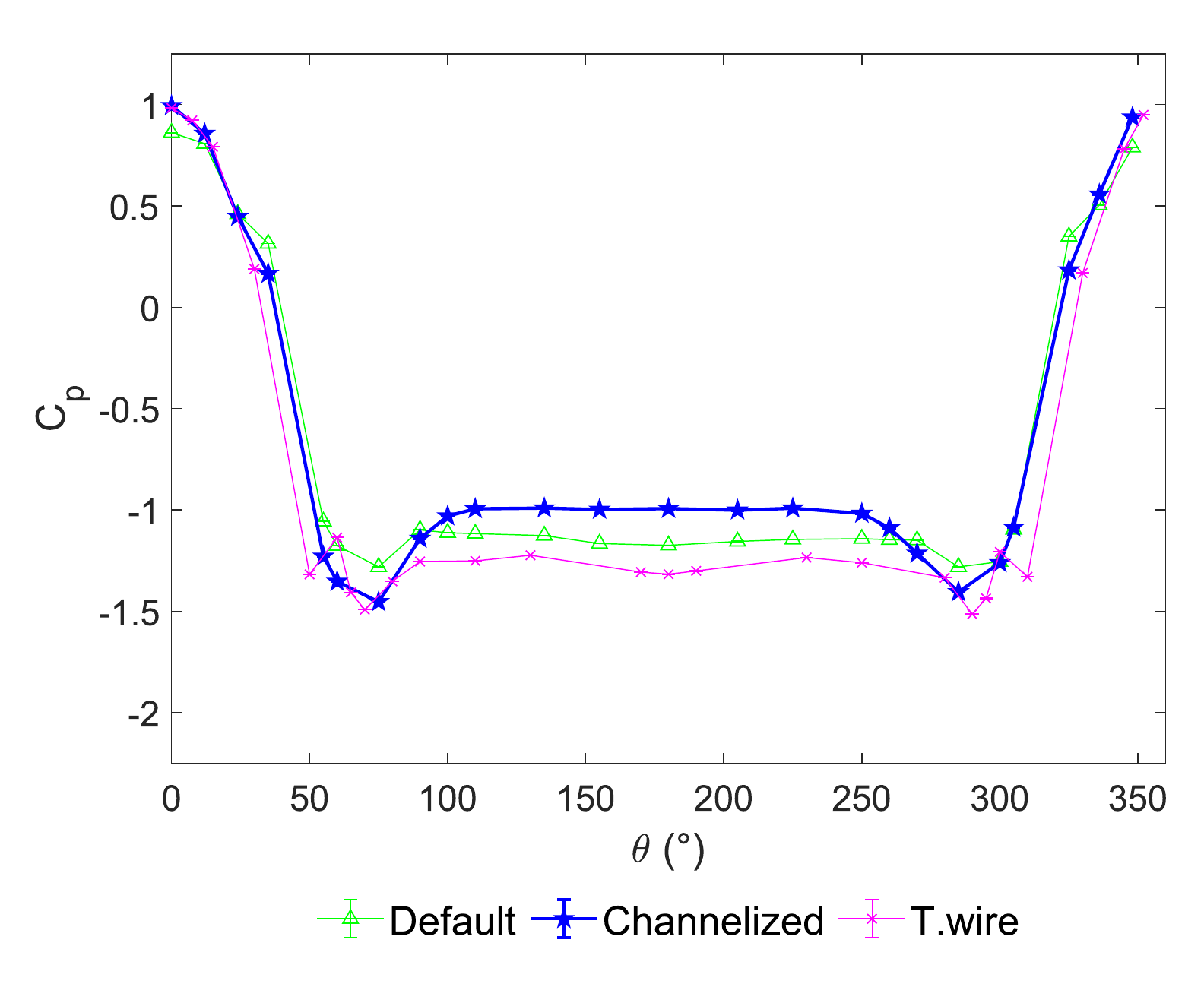}
	\caption{\textbf{\rule[-2pt]{1pt}{13.5pt} \hspace{1pt} The pressure coefficient profiles of the Channelized topology against the Default, and the trip-wire of 0.9\% d/D.} Consistent with the $C_D$ trend, the Channelized topology shows a greater base pressure recovery compared with the trip-wire of 0.9\% d/D at the lowest Re number. At the same time, it does not exhibit the kinks in the $C_p$ profile, otherwise present for the smooth cylinder equipped with tripping devices, strengthening the hypothesis of an alternative working mechanism of boundary layer turbulence promotion, here inferred as distributed passive-jets.}
	\label{Fig:PressureProfileChannelized}
\end{figure}

\begin{figure}[H]
	\centering
	\includegraphics[width=1.00\textwidth]{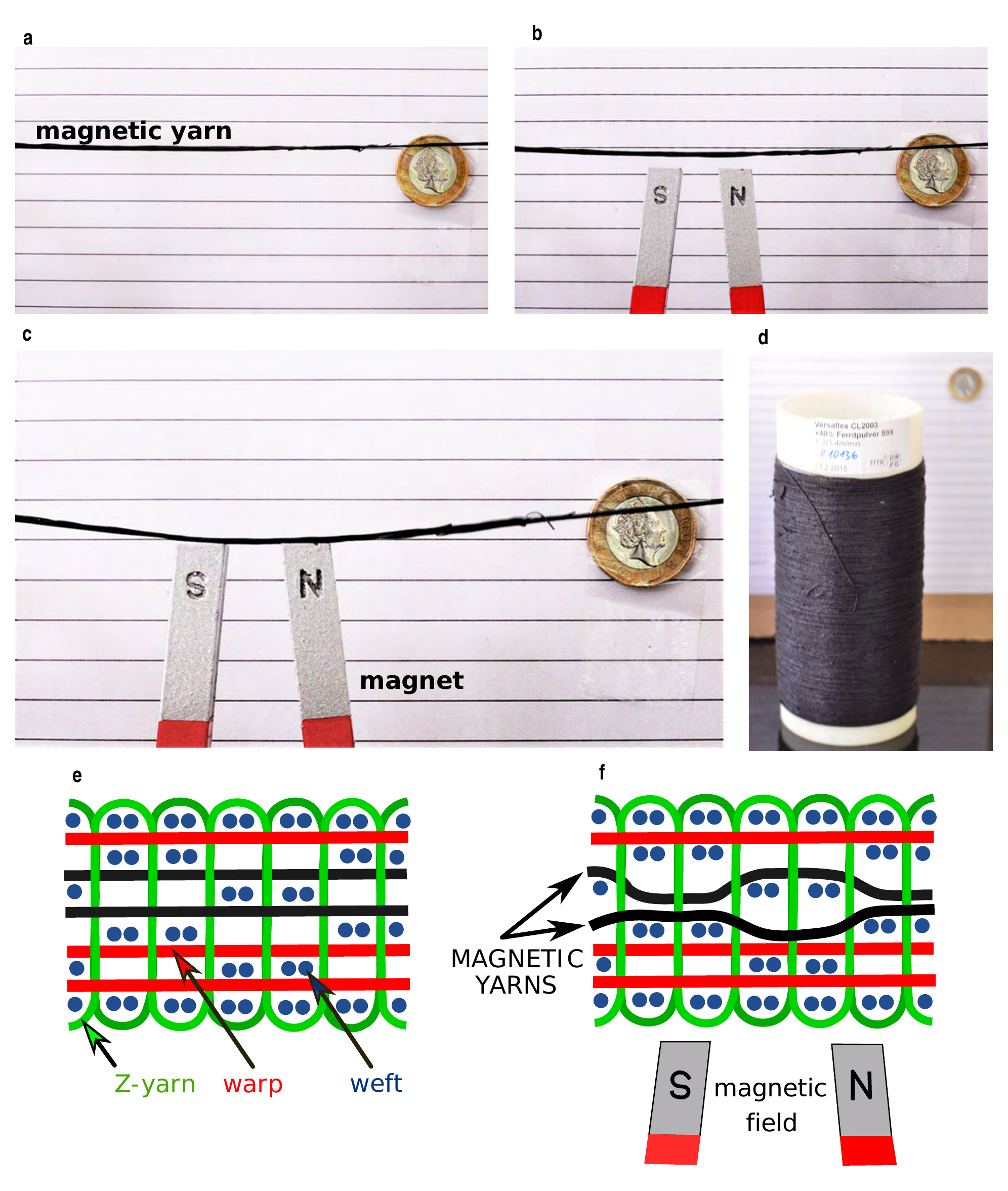}
	\caption{\textbf{\rule[-2pt]{1pt}{13.5pt} \hspace{1pt} The potential for active control of intra-lattice topology using magnetic field.}}
	\label{Fig:MagneticYarns}
\end{figure}

\begin{figure}[H]
	\contcaption{\textbf{\rule[-2pt]{1.5pt}{13.5pt}} \hspace{1pt} \textbf{a}, Magnetic yarn in a neutral state. The yarn material contains 40 wt\% (weight percent) of magnetic Carbonyl-Iron and 60 wt\% of the polymer (VersaflexCL2003X from PolyOne). \textbf{b}, Magnetic field produced by a magnet deflected the yarn. We used traditional Alnico horseshoe magnet with 1 kg pull (75 x 39 x 8 mm). \textbf{c}, Significant deformation of the hybrid polymeric-magnetic yarn was observed. \textbf{d}, A bobbin of the 'magnetic yarn' with 50-micron fibre diameter was provided by M. Schrödner (TITK Rudolstadt) und M. Krautz (IFW Dresden). \textbf{e}, Conceptual design of 3D woven lattice with selected magnetic yarns (shown in black). \textbf{f}, Conceptual deformations of the magnetic yarns under influence of an external magnetic field, which could enable active control of the intra-lattice topologies, and consequently the properties of the intra-lattice and external flows.}
\end{figure}

\section*{Acknowledgements}
This study was funded by the Research Framework of the European Commission under METFOAM Career Integration Grant 631827 with support from program manager Dr Ing. Antonio Cipollaro. The work was also supported by the impact acceleration grant no EP/P511456/1, provided by the Engineering and Physical Science Council (EPSRC) in the UK. Support of Dr Sue Angulatta, a local program manager, is genuinely appreciated. Any opinions, findings, and conclusions expressed in this article are those of the authors and do not necessarily reflect the views of the European Commission nor EPSRC.

We are indebted to Dr Yu Liu from the Southern University of Science and Technology (formerly the University of Surrey) for his initial impulse to study the effect of porous materials on the flow around a circular cylinder experimentally. We are also grateful to Dr Joao Aguiar, who studied open-cell foams under the supervision of Dr Liu, and later with guidance from Dr Dave Birch. We would also like to thank ENFLO lab members at the University of Surrey, for their guidance and help with the testing setups as well as ensuring the highest measurement standards, namely Dr Paul Hayden, Dr Paul Nathan, and Hon. Dr Allan Wells. We also would like to thank the University workshop staff, namely Steve, Lee and Alan for manufacturing of the experimental rig, and Mr Myles who assisted us with 3D printing and cleaning samples from the supporting material. We are also indebted to Dr John Doherty, Dr Matteo Carpentieri, Dr Marco Placidi, and Dr Olaf Marxen for their discussions and observations regarding our study. 

We also would like to thank Prof. Charles Meneveau, Prof. Rajat Mittal and Dr Jung-Hee Seo from the Johns Hopkins University for fruitful discussions about the intra-lattice flow mechanisms. We are grateful to Dr Inaki Caldichoury, Dr Rodrigo Paz and Russell Sims from LSTC (LS-DYNA, CFD solver) for their help with the exploration of modelling techniques and discussions about possible routes to computational optimisation. We would like to thank Prof. Charles-Henri Bruneau and Dr Iraj Mortazavi from the University of Bordeaux in France, for their sharing their articles about computational modelling of external aerodynamics, accounting for the secondary flow via porous inserts ex. We are also indebted to Prof Bodo Ruck from Karlsruhe University of Technology (KIT) for providing a critical appraisal of our work and outlining potential future directions and applications. We would like to thank Dr Judah Goldwasser from the US Office of Naval Research (ONR) for discussions about active control through the use of our material structures.  

We are indebted to Dr Mustafanie M. Yussof for his assistance with drop impact tests to demonstrate crashworthiness of our 3D lattice material structures. We are grateful to M. Schrödner (TITK Rudolstadt) und M. Krautz (IFW Dresden) for supplying us with magnetic yarns, and fruitful discussions about the potential for the use of the magnetic field to control the intra-lattice topologies as achieved in their adaptive filters.

We would like to thank Emily Swinburne, David Robson and Micheal Costa, as well as numerous Brazilian exchange students (Students Without Borders program) for their involvement in wind tunnel testing and manufacturing of 3D lattice material samples as part of their undergraduate and MSc projects.

\newpage
\section*{References}
\bibliography{ArticleCitations}  
\bibliographystyle{naturemag} 
\end{document}